\documentclass[12pt]{article}
\usepackage[latin1]{inputenc}

\usepackage{amsmath}
\usepackage{amsfonts}
\usepackage{amssymb}
\usepackage{graphicx}
\usepackage{geometry}
\usepackage{amssymb,epsfig,subfigure}
\usepackage{hyperref}

\makeatletter
\renewcommand\section{\@startsection {section}{1}{\z@}%
                                 {-3.5ex \@plus -1ex \@minus -.2ex}
                                   {2.3ex \@plus.2ex}%
                                   {\normalfont\large\bfseries}}
\renewcommand\subsection{\@startsection{subsection}{2}{\z@}%
                                   {-3.25ex\@plus -1ex \@minus -.2ex}%
                                     {1.5ex \@plus .2ex}%
                                     {\normalfont\bfseries}}
\renewcommand\subsubsection{\@startsection{subsubsection}{3}{\z@}%
                                   {-3.25ex\@plus -1ex \@minus -.2ex}%
                                     {1.5ex \@plus .2ex}%
                                     {\normalfont\itshape}}
\makeatother

\def\pplogo{\vbox{\kern-\headheight\kern -29pt
\halign{##&##\hfil\cr&{\ppnumber}\cr\rule{0pt}{2.5ex}&\ppdate\cr}}}
\makeatletter
\def\ps@firstpage{\ps@empty \def\@oddhead{\hss\pplogo}%
  \let\@evenhead\@oddhead 
}
\def\maketitle{\par
 \begingroup
 \def\thefootnote{\fnsymbol{footnote}}
 \def\@makefnmark{\hbox{$^{\@thefnmark}$\hss}}
 \if@twocolumn
 \twocolumn[\@maketitle]
 \else \newpage
 \global\@topnum\z@ \@maketitle \fi\thispagestyle{firstpage}\@thanks
 \endgroup
 \setcounter{footnote}{0}
 \let\maketitle\relax
 \let\@maketitle\relax
 \gdef\@thanks{}\gdef\@author{}\gdef\@title{}\let\thanks\relax}
\makeatother

\numberwithin{equation}{section}

\newcommand{\h}[1]{\hat{#1}}

\renewcommand{\dag}{\dagger}

\newcommand{\be}{\begin{equation}}
\newcommand{\bea}{\begin{eqnarray}}
\newcommand{\ee}{\end{equation}}
\newcommand{\eea}{\end{eqnarray}}

\newcommand{\mc}{\mathcal}

\newcommand{\tr}{{\rm tr}}
\renewcommand{\t}{\tilde}
\newcommand{\muphi}{\mu_\phi}

\begin{document}

\setcounter{page}0
\def\ppnumber{\vbox{\baselineskip14pt
}}
\def\ppdate{\footnotesize{SLAC-PUB-14293, NSF-KITP-10-138, KCL-MTH-10-13}} \date{}

\author{Sakura Sch\"afer-Nameki$^{1\&3}$, Carlos Tamarit$^{1}$, Gonzalo Torroba$^{2}$\\
[7mm]
{\normalsize $^1$ \it Kavli Institute for Theoretical Physics}\\
{\normalsize \it University of California, Santa Barbara, CA 93106, USA}\\
{\normalsize $^2$ \it SLAC National Accelerator Laboratory, }\\
{\normalsize \it Stanford University, Stanford, CA 94309, USA}\\
{\normalsize $^3$ \it Department of Mathematics, King's College London }\\
{\normalsize \it The Strand, WC2R 2LS, London, England}\\
[3mm]
{\tt \footnotesize ss299 at  theory.caltech.edu, tamarit at kitp.ucsb.edu, torrobag at slac.stanford.edu }
}

\bigskip
\title{\bf  Naturalness from runaways in \\ direct mediation
\vskip 0.5cm}
\maketitle

\begin{abstract} \normalsize
\noindent Postulating that the NMSSM singlet is a meson of a microscopic confining theory opens up new model-building possibilities. Based on this, we construct calculable models of direct mediation that solve the $\mu/B_\mu$ problem and simultaneously lead to realistic phenomenology. The singlet that couples to the Higgs fields develops a runaway produced by soft interactions, then stabilized by a small superpotential perturbation. The mechanism is first realized in an O'Raifeartaigh model of direct gauge mediation with metastable supersymmetry breaking. Focusing then on the microscopic theory, we argue that super QCD with massless and massive flavors in the free magnetic phase gives rise to this dynamics in the infrared. A deformation of the SQCD superpotential leads to large spontaneous R-symmetry breaking, gaugino masses naturally at the scale of the Higgs mass parameters, and absence of CP violating phases.
\end{abstract}
\bigskip
\newpage

\tableofcontents

\vskip 1cm

\section{Introduction and overview}\label{sec:intro}

A central goal of models of gauge mediation is to break $SU(2) \times U(1)$ naturally and produce realistic Higgs masses. A very important challenge is the $\mu/B_\mu$ problem~\cite{Dvali:1996cu}. Starting from the supersymmetric and soft mass terms
\be\label{eq:Lagr1}
W = \mu H_u H_d\;,\;V_{soft}= m_{H_u}^2 |H_u|^2+ m_{H_d}^2 |H_d|^2 + (B_\mu H_u H_d+c.c.)\,,
\ee
supersymmetry breaking mechanisms typically generate $\mu$ and $B_\mu$ at the same loop order,
\be\label{eq:wrong}
B_\mu \sim 16 \pi^2 \mu^2 \gg \mu^2\,.
\ee
This leads to unacceptable phenomenology, with no natural solution to the EWSB conditions.

Rather than viewing the $\mu/B_\mu$ problem as a drawback of gauge-mediation,  it should be considered as an important hint into a more fundamental theory and as a useful guide for building new models. Successful solutions to $\mu/B_\mu$ have often lead to new model-building avenues.\footnote{Some of these include: the dynamical relaxation mechanism of~\cite{Dvali:1996cu}, couplings between the Higgs fields and hidden sector singlets/doublets (see~\cite{Komargodski:2008ax} for a general classification and references), models with a hierarchy $m_{H_d}^2 \gg m_{H_u}^2$ and compositeness~\cite{Csaki:2008sr,SchaferNameki:2010iz}, strong dynamics~\cite{Roy:2007nz,Murayama:2007ge} and connections to R-symmetry~\cite{Hall:2002up,Dine:2009swa}. Other recent developments include~\cite{Delgado:2007rz}-\cite{Evans:2010ru}. See e.g.~\cite{Giudice:1998bp} for further references.}

The core of the problem is the relation (\ref{eq:wrong}), and various mechanisms have been devised where $\mu$ is generated at one loop, while $B_\mu$ appears starting from two loops. Nevertheless, in many cases the models accomplishing this are somewhat contrived, requiring various levels of interactions and messengers.  In trying to construct simpler models that address $\mu/B_\mu$ and have a realistic phenomenology, we are naturally led to consider \textit{direct mediation} of supersymmetry breaking (see~\cite{Luty:1998vr}). The goal of this work is to construct realistic models of direct mediation that solve $\mu/B_\mu$ dynamically.

The analysis will be carried out in the context of the NMSSM~\cite{Ellis:1988er}, where a new singlet $S$ couples to the Higgs fields through a superpotential term.
The NMSSM provides one of the simplest frameworks to address $\mu/B_\mu$, in which the $\mu$ term arises dynamically from the expectation value of the singlet. In order to avoid a new mass scale for $S$, the superpotential is traditionally taken to be
\be\label{eq:introW}
W\supset\lambda\,S H_u H_d+ S^3\,.
\ee 
The simplest realizations of this idea in gauge mediation do not produce phenomenologically acceptable spectra, and new dynamics is needed to generate a tachyonic mass for $S$ or enhance some of the soft terms \cite{deGouvea:1997cx}, which demands the introduction of additional fields or couplings.\footnote{Known ways to generate a runaway include: coupling the singlet to additional vector-like matter multiplets with indirect couplings to the susy breaking fields \cite{Dine:1993yw,Dine:1994vc,Evans:2010ru}, coupling the singlet directly to messengers or other singlets \cite{Dvali:1996cu,Delgado:2007rz,Dine:1994vc,Giudice:1997ni,Dine:2006xt,Dine:2007dz}, or coupling the singlet to the susy spurion \cite{Dine:1994vc}.
See also~\cite{Komargodski:2008ax} for a treatment of the subject in the framework of general gauge mediation. } 

We will see that direct mediation mechanisms based on a confining supersymmetric gauge theory can realize the dynamics necessary to achieve phenomenologically viable realizations of the NMSSM, without the need to introduce fields outside the sector of supersymmetry breaking. In this context, it will be natural to consider models where $S$ is given a supersymmetric mass term, which are in general not analyzed in the literature --partly of course because of the additional mass scale. There are in fact various reasons for investigating this possibility. First, the mass scale can be made naturally small as in~\cite{Dine:2006gm,Essig:2007xk}. Next, a superpotential of the form (\ref{eq:introW}) but with a quadratic term for $S$ leads to dimension five operators $(H_u H_d)^2$. These operators encode leading corrections from new physics potentially at the TeV scale~\cite{Dine:2007xi}\footnote{We thank M. Dine for discussions on this point.}, and thus it is interesting to consider them in an effective description beyond the MSSM. In our approach, the main motivation comes from strong gauge dynamics, as we explain next.


Our chief proposal is that the NMSSM singlet can secretly be a composite field of a microscopic confining gauge theory, and the main result  will be that this gauge theory can  break supersymmetry dynamically and simultaneously solve $\mu/B_\mu$, leading to quite economical models. Moreover, these will be models of direct mediation, without additional messenger sectors. The simplest realization will be in SQCD with massless and massive flavors $(Q, \t Q)$, in the free magnetic phase.

In view of this, the motivation to consider a mass term for $S$ in the superpotential is the following. If the singlet is a meson, $S= (Q \t Q)$, then the superpotential coupling to the Higgs fields is in fact a quartic operator in the microscopic theory,
\be
W \supset  \frac{1}{\Lambda_0}(Q \t Q) H_u H_d\,,
\ee
where $\Lambda_0$ is some scale higher than the confining scale $\Lambda$. In this case, we should also consider the quartic operator $(Q \t Q)^2$, and this leads to a mass term for $S$ in the confined theory, with the mass generated by $\Lambda^2/\Lambda_0$. Therefore, from the point of view of a microscopic confining theory, it is quite natural to also have a mass for $S$. Both are naturally small, coming from irrelevant perturbations. The cubic coupling would on the other hand arise from a dimension 6 operator $(Q \t Q)^3$.

Our approach will proceed in three steps. First, we present in \S \ref{sec:sol} a mechanism to achieve natural $\mu$ and $B_\mu$ in the NMSSM in which the singlet $S$ develops a runaway from soft supersymmetry breaking, which is eventually stabilized by a small superpotential perturbation 
\be
W \supset t S^k\,.
\ee
This mechanism is analyzed in a general effective theory that includes only the singlet and Higgs fields, and the EWSB conditions and spectrum are studied keeping the soft potential and $k$ arbitrary. Next in \S \ref{sec:OR} we construct an O'Raifeartaigh model of direct mediation that realizes the previous effective theory. Different stabilization mechanisms are discussed, corresponding to both $k=2$ and $k=3$. 

In the third step, it is argued that SQCD in the free magnetic phase with massive and massless flavors $(Q, \t Q)$ provides a natural dynamical realization for our macroscopic model; here $S$ and the O'Raifeartaigh supersymmetry breaking field arise as composite mesons $(Q \t Q)$. This will be discussed in \S \ref{sec:SQCD}. One of the results is that in the theory with $k=2$, $\mu$ and $B_\mu$ are generated at the same order, yet $\mu^2 \sim B_\mu$. This provides an interesting alternative to mechanisms where these arise at different loop orders. The phenomenology of this model is explored in \S \ref{sec:k2}.

\section{Solving $\mu/B_\mu$ along a runaway direction}\label{sec:sol}

In this section we analyze a general mechanism whereby the NMSSM singlet $S$ that couples to the Higgs fields develops a runaway, which is eventually stabilized by a small superpotential perturbation $W \supset t S^k$, leading to phenomenologically acceptable values for $\mu$ and $B_\mu$. This works as follows: in the limit $t \to 0$, due to the runaway we would have $S \to \infty$; next turning on a finite but small $t$ should give a large expectation value for $S$, but $B_\mu/\mu^2$ can be suppressed by powers of the small parameter. The smallness of $t$ will be explained dynamically in \S \ref{sec:SQCD}.

We work in an effective theory containing only $S$ and the Higgs fields, with superpotential
\be\label{eq:Weff}
W=\lambda\,S H_u H_d+ \frac{t}{k}\,S^k
\ee
plus a soft potential $V_{soft}(S)$ that produces the runaway for $S$. The superpotential is renormalizable for $k \le 3$, although the analysis is performed for general $k \ge 2$.\footnote{ We do require $k > 1$ so that the EWSB conditions are not radically modified by the superpotential interactions. This can be enforced by the symmetries described below.} Soft A-terms $L \supset A_\lambda S H_u H_d$ are in general negligible in gauge mediation with low scale supersymmetry breaking and will be not be considered here. In the next section we will obtain this effective theory from an O'Raifeartaigh model. Then we will show that this mechanism arises very naturally in SQCD in the free magnetic phase.

The superpotential (\ref{eq:Weff}) has a discrete $\mathbb Z_k$ symmetry that forbids a tree-level $\mu$ term and additional interactions for $S$.
The continuous abelian symmetries of the theory are
\begin{center}
\begin{tabular}{c|ccc}
&$U(1)_V$&$U(1)_A$&$U(1)_R$\\
\hline
$S$ & $0$ & $-2$ &  $2/k$\\
$H_u$ & $1$ & $1$ &  $1-1/k$\\
$H_d$ & $-1$ & $1$ & $1-1/k$\\
$t$ & $0$ & $2k$ &  $0$
\end{tabular}
\end{center}
The interaction parameter $t$ breaks the axial $U(1)_A$; this is then only an approximate symmetry but is still useful since $t S^k$ will be a small perturbation. The R-symmetry $U(1)_R$ is exact at tree-level and will play an important role in what follows. Also, some of these symmetries will become anomalous once the model is realized dynamically in a gauge theory. For $k \le 3$, (\ref{eq:Weff}) is the most general renormalizable polynomial superpotential allowed by the above exact symmetries.\footnote{The operators $W \supset (H_u H_d)^{k/(k-1)}$ are also consistent with symmetries and can be generated for instance by integrating out $S$ supersymmetrically. In our case they are either irrelevant or non-polynomial; they do not modify our conclusions and will be set to zero in the following.}

As will be discussed in \S \ref{sec:OR}, the nonsupersymmetric contribution $V_{soft}(S)$ that is responsible for the runaway can be obtained by integrating out heavy fields in a theory with spontaneous supersymmetry breaking. We will assume that these interactions respect $U(1)_R$, so that $V_{soft}$ can only depend on $|S|$. The soft potential depends sensitively on the details of the supersymmetry breaking model. In a limit of small supersymmetry breaking (compared to the supersymmetric masses), $V_{soft}$ can have a logarithmic behavior from wave-function renormalization. On the other hand, in models where the supersymmetry breaking scale is comparable to the supersymmetric masses we would expect a polynomial dependence. In fact, both limits will be realized in \S \ref{sec:OR}. However, in this section we keep the analysis general and do not restrict to any particular potential.

Since the expectation value of $S$ is controlled by the inverse of the small parameter $t$, our class of models will have $\langle S \rangle \gg \langle H \rangle$. At this order it is consistent to ignore the Higgs self-interactions. The EWSB conditions will be discussed in more detail in \S \ref{subsec:singlet}. Then the potential reads
\be\label{eq:Veff}
V = |\lambda|^2|S|^2 \left(|H_u|^2+|H_d|^2 \right) + |\lambda H_u H_d + t S^{k-1}|^2 + V_{soft}(|S|)\,.
\ee
First, setting $V_{soft}=0$, there are supersymmetric vacua at $S=H_u = H_d=0$. On the other hand, turning off the superpotential interactions, $V_{soft}$ gives a runaway $S \to \infty$. In order to produce a minimum at nonzero $S$, $V_{soft}$ has to destabilize the origin but the total potential should have positive slope at large values of $S$,
\bea\label{eq:conds}
V'_{soft}& <&0\;\;\textrm {at}\;\; S=0 \nonumber\\
\frac{V''_{soft}}{|S|^{2k-4}} &\to& 0\;{\rm for}\;S \gg 0\,.
\eea
This grants the existence of a minimum at some intermediate point $S=S_0$. A microscopic model that realizes our mechanism has to verify these conditions. (Here and in what follows, `primes' denote derivatives with respect to $|S|$.)

The minimum $S=S_0$ is at
\be\label{eq:S0}
|S_0|^{2k-3}= \frac{1}{2(k-1)}\,\left|\frac{V'_{soft}}{t^2} \right|\;\;,\;\;H_u=H_d=0\,,
\ee
where $V'_{soft}$ is evaluated at $S=S_0$. The runaway is recovered for $t \to 0$. The expectation value of $S$ spontaneously breaks the $U(1)_R$ symmetry; the associated Nambu-Goldstone boson is its phase $\phi_S$,
\be
S_0 = |S_0|\,e^{i \phi_S}\,.
\ee
The stabilization of the R-axion will be addressed in the realistic model below. Let us now discuss the generation of $\mu$ and $B_\mu$.

\subsection{Natural $\mu$ and $B_\mu$}\label{subsec:natural}

Combining (\ref{eq:S0}) with Eqs.~(\ref{eq:Weff}), (\ref{eq:Veff}), gives
\bea\label{eq:muBmu}
\mu & = & \lambda S_0 = \lambda e^{i \phi_S} \left(\frac{1}{2(k-1)} \left| \frac{V_{soft}'}{t^2}\right| \right)^{1/(2k-3)}\nonumber\\
B_\mu & =& \lambda (t S_0^{k-1})^* = \lambda t^* e^{-i(k-1)\phi_S} \left(\frac{1}{2(k-1)} \left| \frac{V_{soft}'}{t^2}\right| \right)^{(k-1)/(2k-3)}\,,
\eea
where $V'_{soft}$ is evaluated at $S_0$. $B_\mu$ arises from the cross-coupling between $H_u H_d$ and $S^{k-1}$ in the F-term $W_S$ and has an additional $t$ suppression.

Next, having consistent electro-weak symmetry breaking and small fine-tuning, requires 
\be\label{eq:natural2}
\mu^2 \sim B_\mu\;\Rightarrow\;|S_0|^{k-3} \sim \frac{|\lambda|}{|t|}\,.
\ee
Imposing this condition relates $V_{soft}'$ to the superpotential parameters,
\be\label{eq:natural}
\left| \frac{V'_{soft}}{t^2}\right|^{\frac{k-3}{2k-3}} \sim \frac{|\lambda|}{|t|}\,.
\ee
In the microscopic realizations below, this will constraint the scale of supersymmetry breaking at which $V_{soft}$ is generated. We find, for $k \neq 3$,
\be
B_\mu \sim \mu^2 \sim \left( \lambda^{4-2k} t^2 \right)^{1/(3-k)}
\ee
(ignoring phases), while the result for $k=3$ gives $t \sim \lambda$ and
\be
B_\mu \sim \mu^2 \sim \left( \lambda V'_{soft}\right)^{2/3}\,.
\ee

Therefore, coupling the MSSM Higgs fields to a singlet that has a nonsupersymmetric runaway stabilized by a small superpotential perturbation $t$, leads to a simple solution of the $\mu/B_\mu$ problem. $\mu^2$ is generated by balancing the runaway against the $t$-perturbation and is naturally larger than $B_\mu$, which has an additional {\textit{suppression}} from $t$. Indeed, keeping all the other parameters fixed,
$$
\frac{B_\mu}{\mu^2} \propto\,t^{1/(2k-3)} \,\to\,0\;\;{\rm when}\;\;t \to 0
$$
which grants that there exist values of the couplings for which $B_\mu \sim \mu^2$ even though both are generated at the same order (tree-level in the effective theory containing only the singlet and Higgs fields).

Note that the naturalness criterion for $k=3$, which implies $t\sim\lambda$, introduces a coincidence problem between dimensionless couplings. This should be explained by the UV completion of the effective model. This is challenging in the specific realizations that we investigate in this work, where the singlet is a composite meson of a SQCD theory and then $t$ and $\lambda$ come from operators of different ultraviolet dimensions. The case $k=2$ will be shown to be more natural. Indeed, Eq.~\eqref{eq:natural2} implies $t\sim\mu$, a result which will be related to the mechanism for providing gaugino masses. In any case, our effective analysis and conclusions here are general, and more natural UV completions for $k=3$ may exist.

In a given model, the superpotential of Eq.~\eqref{eq:Weff} will be supplemented by further interactions, which in general involve an additional susy spurion $X$ that breaks supersymmetry with a larger $F$-term, $F_X\gg F_S$. It is important to make sure that this does not generate couplings of the form
\begin{align}
 \int\!d^4\theta X^\dagger S,\, \int\!d^2\theta S X^2.
\label{eq:Stadpoles}
\end{align}
The first term would generate kinetic mixing that would typically yield an unacceptably large $B_\mu$, while the second term would generate undesired contributions to $\mu$. Symmetries can guarantee that these terms are suppressed or not generated, as we will see in the models of the following sections. 

\subsection{EWSB and mass spectrum}\label{subsec:singlet}

So far, our analysis has involved only $S$, ignoring contributions from the Higgs fields. This is justified self-consistently because $\langle S \rangle \gg \langle H_u \rangle, \langle H_d \rangle$. On the other hand, considering small fluctuations around the vacuum reveals that $|S|$ has a mass
\be
m_s \sim \frac{B_\mu}{\mu}\,.
\ee
Thus, imposing $B_\mu \sim \mu^2$ implies that the singlet fluctuations do not decouple from the low energy theory. Moreover, there are mass-mixings between singlet and the physical Higgs that cannot be ignored. Therefore the analysis of quadratic fluctuations should be done including $S$ and both Higgses.

The spectrum depends crucially on the CP properties of the vacuum. CP-violating phases have to be avoided in order to obtain a realistic phenomenology. This will be shown to be the case in our models below; in particular, the phase $\phi_S$ can be consistently stabilized at the origin. Accordingly, in what follows the expectation values and superpotential parameters are chosen to be real.  The tree-level equations of motion evaluated at the vacuum
$$
\langle S\rangle=\frac{\mu}{\lambda}\;,\;\langle H_u\rangle=v_u\;,\;\langle H_d\rangle=v_d
$$ 
are
\begin{align}
\nonumber\frac{1}{2}\left.\frac{dV_{\rm soft}(|S|)}{d|S|}\right|_{S_0}&=\lambda v^2\sin2\beta(k-1)\frac{B_\mu}{2\mu}-(k-1)\frac{B^2_\mu}{\lambda\mu}-\lambda\mu v^2,\\
 \label{eq:EWSB}B_\mu&=\frac{\sin 2\beta}{2}(m^2_{H_u}+m^2_{H_d}+2\mu^2+\lambda^2 v^2),\\
\nonumber\mu^2&=-\frac{M^2_Z}{2}+\frac{m^2_{H_d}-m^2_{H_u}\tan^2\beta}{\tan^2\beta-1},
\end{align}
where we have used $B_\mu=\lambda t \langle S\rangle^{k-1}$. 

With the help of the above equations of motion, one can obtain the mass matrices for the fluctuations of the fields $S,H_u,H_d$. For a $CP$-invariant vacuum, the real and imaginary parts do not mix. As usual, there is a Goldstone mode in the imaginary sector, given by 
$$
G_0=-\sin\beta\,{\rm Im}H^0_u+\cos\beta\,{\rm Im}H^0_d\,,
$$
which is eaten by the Higgs mechanism. The scalar and pseudoscalar states orthogonal to this mode are grouped into
\be
\phi_S^\intercal={\rm Re}[ H^0_u,\,\,H^0_d,\,\, S],\quad\phi_P^\intercal={\rm Im}[ \cos\beta H^0_u+\sin\beta H^0_d,\,\, S]\,.
\ee
All the fields are understood as fluctuations around the vacuum expectation values. The corresponding mass-matrices are found to be
\begin{align}
\nonumber&M^2_S=\\
\nonumber&\left[
\begin{array}{ccc}
 \!B_\mu \cot\beta+m^2_Z\sin^2\beta, & -B_\mu+\sin2\beta(\lambda^2v^2-\frac{1}{2}m^2_Z)  , & v\lambda(2\mu \sin\beta\!-\!\cos\beta(k\!-\!1)\frac{B_\mu}{\mu})\\
\!-B_\mu+\sin2\beta(\lambda^2v^2-\frac{1}{2}m^2_Z) ,& B_\mu \tan\beta+m^2_Z\cos^2\beta,& v\lambda(2\mu \cos\beta\!-\!\sin\beta(k\!-\!1)\frac{B_\mu}{\mu} )\\
v\lambda(2\mu \sin\beta\!-\!\cos\beta(k\!-\!1)\frac{B_\mu}{\mu}), & v\lambda(2\mu \cos\beta\!-\!\sin\beta(k\!-\!1)\frac{B_\mu}{\mu} ),& m^2_{S}
\end{array}
\right]\!,\\
\nonumber&M^2_P=\left[
\begin{array}{cc}
\frac{2B_\mu}{\sin2\beta} , &  -\frac{B_\mu}{\mu}(k-1)\lambda v \\
 -\frac{B_\mu}{\mu}(k-1)\lambda v, & v^2\sin2\beta\lambda^2(k-1)^2\frac{B_\mu}{2\mu^2}
\end{array}
\right],\\
&m^2_{S}=2(k-1)(k-2)\frac{B^2_\mu}{\mu^2}+v^2\sin 2\beta\lambda^2(k-1)(3-k)\frac{B_\mu}{2\mu^2}+\frac{2\mu^2}{\lambda^2}\frac{d^2V_{\rm soft}}{d(|S|^2)^2}.
\label{eq:masses}
\end{align}

In the CP-even sector, the mixing between the Higgs fields and the singlet affects the value of the lightest CP-even neutral particle. It turns out 
that this mixing produces typically a decrease in the lightest mass with respect to its MSSM value, which can be estimated using perturbation theory around vacuum configurations satisfying the naturalness criterion $B_\mu\sim\mu^2\gg m^2_Z,m^2_h$  as
\be
\delta m_h^2  \approx \lambda^2 \sin^2(2\beta) v^2-\frac{1}{2}\lambda^2 v^2   \frac{\left( (k-1)\sin(2 \beta)(B_\mu/\mu^2)- 2\right)^2}{ (k-1 )(k-2) (B_\mu/\mu^2)^2 +1}\,.
\ee
More details are given in Appendix \ref{app:mass-shiftSSN}. This typically yields a shift of the order of tens of GeV for $\lambda\sim0.1$ (note that this is only a tree-level result, and radiative corrections will be significant, as is known for the MSSM). 

The mass of the CP-even singlet-like eigenvalue can be estimated from $m^2_S$ in Eq.~\eqref{eq:masses}. Imposing the naturalness requirement $B_\mu\sim\mu^2$,
and using Eq.~\eqref{eq:EWSB}, one has
 \be
V'_{\rm soft}\sim\frac{B_\mu}{\lambda\mu}\sim\frac{\mu}{\lambda}.
\ee
In the vacuum, $S=\frac{\mu}{\lambda}$, from which one can estimate 
\be
\frac{d^2 V_{\rm soft}(|S|)}{d(|S|^2)^2}\sim \frac{1}{S^3}V'_{\rm soft}\sim \lambda^2,
\ee 
so that $m^2_S$ in Eq.~\eqref{eq:masses} is of order $\mu^2$, as anticipated. The impossibility of decoupling the singlet from the low energy theory is a consequence of the runaway mechanism that we have employed to generate the Higgs soft parameters. Similar effects were already observed, for instance, in~\cite{Ellis:1988er}. If the higgsinos and physical Higgs bosons have masses of order 1 TeV, there will also be a light singlet $s$ at a similar energy scale; the existence of this extra particle at the TeV scale may have interesting consequences on the low energy phenomenology. Via its mixings with the light Higgs it may provide a window into some of the properties of the hidden sector underlying the effective model (see \S \S \ref{sec:OR}, \ref{sec:SQCD}).

Finally, the CP-odd mass matrix $M^2_P$ contains, as expected, a massless eigenvalue corresponding to the R-axion. This field will acquire a mass in our model of direct mediation, in connection with the mechanism that generates realistic gaugino masses.

\section{Macroscopic model of direct mediation}\label{sec:OR}

In this section we construct a model that breaks supersymmetry spontaneously and that gives rise to the theory of \S \ref{sec:sol}. The calculable description of the supersymmetry breaking dynamics is given, as usual, in terms of the O'Raifeartaigh theory. The mechanism that we will use to transmit supersymmetry breaking to the visible sector will be through {\textit {direct gauge mediation}}. Here the Standard Model gauge bosons couple directly to the supersymmetry breaking sector; messengers are not added in as a separate module, but rather are an integral part of the dynamics responsible for supersymmetry breaking.

Our choice for the O'Raifeartaigh model of direct mediation corresponds to the `macroscopic model' of Intriligator, Seiberg and Shih (ISS)~\cite{Intriligator:2006dd}:
\be\label{eq:OR}
W_{O'R}= -h \mu_2^2\, \tr\,X + h\,\tr(X \rho \t \rho) + h \mu_1\,\tr(\rho \t Z + \t \rho Z)
\ee
where all the superpotential parameters are taken to be real and $\mu_1 > \mu_2$. Here $X$ is an $N \times N$ matrix, while the rest of the fields are $N$-component vectors. In fact, choosing this model will also allow us to show how our theory including the runaway arises dynamically from SQCD in the free magnetic phase. This will be the subject of \S \ref{sec:SQCD}. Before adding in the singlet and explaining how to produce the desired runaway, we briefly review how supersymmetry is broken
in the theory with superpotential (\ref{eq:OR}).

\subsection{Review of the O'Raifeartaigh model of supersymmetry breaking}\label{subsec:review}

Supersymmetry is broken because the F-terms of $Z$ and $\t Z$ force $\rho = \t \rho = 0$ but then $W_X=0$ cannot be satisfied. For $\mu_1 \ge \mu_2$, the supersymmetry breaking vacuum has
\be
W_{X} = - h \mu_2^2\,\mathbf 1_{N \times N}\;\;,\;\;V_0 =N\,(h \mu_2^2)^2\,.
\ee
The field $X$ is a ``pseudo-modulus'': it is massless at tree-level but will be lifted by quantum effects. The rest of the fields are massive, with supersymmetric masses and nonsupersymmetric splittings
\be
M = h \mu_1\;,\;F^{1/2}=h \mu_2\,;
\ee
at the vacuum, $\rho = \t \rho = Z = \t Z =0$. 

Integrating out the massive fields produces a one-loop Coleman-Weinberg (CW) potential
\be
V_{CW} = \frac{1}{64 \pi^2}\,{\rm Str}\,\mc M^4\,\log\,\frac{\mc M^2}{(h \mu_1)^2}
\ee
where $\mc M$ stands the bosonic and fermionic mass matrices for the massive fields, keeping $X$ as a background superfield. For simplicity, the UV cutoff has been chosen of order $h \mu_1$. These quantum effects stabilize the pseudo-modulus at the origin~\cite{Intriligator:2006dd}
\be\label{eq:CW}
V_{CW} = m_{CW}^2\, \tr(X^\dag X) + \mc O (|X|^4)\;,\;\;m_{CW}^2 \approx  \frac{h^2}{16 \pi^2} \,\left(\frac{h \mu_2^2}{\mu_1} \right)^2,
\ee
(where the superpotential parameters were chosen to be real). Therefore $X$ obtains a positive mass squared proportional to $F/M$ times a loop factor. 

The model (\ref{eq:OR}) has an $SU(N)$ global symmetry with $\rho$ and $Z$ transforming as fundamentals, $\t \rho$ and $\t Z$ as antifundamentals, and $X$ decomposing into an adjoint of $SU(N)$ plus a singlet $\tr\,X$. In the model of direct mediation, a subgroup 
$$
SU(5)_{SM} \subset SU(N)
$$
is weakly gauged and identified with the SM gauge group. Therefore, the fields $(\rho, Z)$ simultaneously participate in the breaking of supersymmetry and mediate these effects to the visible sector.

In this model of direct mediation, $\rho$ and $Z$ give 2-loop gauge mediated sfermion masses
\be\label{eq:mGM}
m_{GM}^2 \sim \left( \frac{g_{SM}^2}{16 \pi^2}\right)^2\,\left(\frac{h \mu_2}{\mu_1} \right)^2\,.
\ee
At this stage, Majorana gaugino masses are forbidden by a $U(1)_R$ symmetry under which $R(X)=2$. We will discuss how to obtain realistic gaugino masses and present further model-building applications in \S \ref{sec:SQCD}.

\subsection{Generating the runaway}\label{subsec:runaway}

To generate the runaway, the singlet $S$ is coupled to a new set of fields $(\rho_0, \t \rho_0, Z_0, \t Z_0)$ with a structure analogous to (\ref{eq:OR}),
\be\label{eq:Wsinglet}
W_{singlet}=h\,S \rho_0 \t \rho_0+ h \mu_1\,(\rho_0 \t Z_0 + \t \rho_0 Z_0)\,.
\ee
The difference with $W_{O'R}$ is that $S$ does not have a linear term, and these new fields are singlets under the $SU(N)$ flavor symmetry. Finally, we add `link fields' $L$ and $\t L$ that couple both sets of messengers,
\be
W_{link} = h\,\tr (\rho \t L) \t \rho_0 + h \,\tr(\t \rho L) \rho_0\,.
\label{eq:Wlink}
\ee
The full superpotential of the supersymmetric theory is
\be\label{eq:Wtotal}
W = W_{O'R} + W_{singlet} + W_{link}+\frac{t}{k} S^k\,.
\ee
We will show in \S \ref{sec:SQCD} how this structure arises very naturally from SQCD.

The continuous abelian symmetries are
\begin{center}
\begin{tabular}{c|cccc}
&$U(1)_V$&$U(1)_A$&$U(1)_A'$&$U(1)_R$\\
\hline
$X$ & $0$ & $-2$ & $0$ & $2$\\
$Z$ & $1$ & $-1$ & $0$ & $2$\\
$\t Z$ & $-1$ & $-1$ & $0$ & $2$\\
$\rho$ & $1$ & $1$ & $0$& $0$ \\
$\t \rho$ & $-1$ & $1$ & $0$ & $0$\\
\hline
$S$ & $0$ & $0$ & $-2$ & $2/k$\\
$Z_0$ & $1$ & $0$ & $-1$ & $1+1/k$\\
$\t Z_0$ & $-1$ & $0$ & $-1$ & $1+1/k$\\
$\rho_0$ & $1$ & $0$ & $1$& $1-1/k$ \\
$\t \rho_0$ & $-1$ & $0$ & $1$ & $1-1/k$\\
\hline
$L$ & $0$ & $-1$ & $-1$ & $1+1/k$\\
$\t L$ & $0$ & $-1$ & $-1$ & $1+1/k$\\
\end{tabular}
\end{center}
The axial symmetries are broken by $\mu_2^2$ and $t$, respectively. There is also a discrete $\mathbb Z_2$ charge conjugation symmetry, interchanging $\rho \to \t \rho$ etc., and a discrete $\mathbb Z_k$ symmetry under which only $S$ and the fields coupling to it are charged. Note that the exact symmetries forbid the dangerous couplings of Eq.~\eqref{eq:Stadpoles}. These symmetries can be used to enforce the form of the superpotential.

Let us discuss the vacuum structure of (\ref{eq:Wtotal}). The addition of $W_{singlet}+W_{link}$ does not modify the supersymmetry breaking vacuum of \S \ref{subsec:review}. The extra set of messengers is stabilized at the origin
\be
\rho_0 = \t \rho_0 = Z_0 = \t Z_0 = 0\,,
\ee
with masses
\be
m_{\pm}^2= (h \mu_1)^2+ \frac{1}{2} (hS)^2 \pm \frac{1}{2} (hS)\,\sqrt{(hS)^2 + 4 (h\mu_1)^2}\,.
\ee
As before, $X$ is the pseudo-modulus breaking supersymmetry. On the other hand, $S$, $L$ and $\t L$ are pseudomoduli without F-terms. The key point for producing the runaway along $S$ is that this field only couples to $(\rho_0, \t \rho_0)$,
which are supersymmetric at tree-level in the limit $t\rightarrow0$. Therefore in this limit there is no one-loop mass -as was first found in ref.~\cite{Franco:2006es} in the context of SQCD with massless and massive flavours- while two-loop effects enter with a sign opposite to (\ref{eq:CW}) and destabilize the origin of $S$. This was studied  by~\cite{Giveon:2008wp} using the two-loop results of~\cite{Martin:2001vx}.

The two-loop potential for $S$ for $t=0$ can be written as
\be\label{eq:VsoftS}
V_{soft}(|S|)= -N\,\left( \frac{h^2}{16 \pi^2}\right)^2\,(h\mu_2^2)^2\,\hat V\left(\frac{|S|}{\mu_1}\right)
\ee
where the loop factor and supersymmetry breaking scale have been made explicit, and $\hat V$ is a dimensionless function depending only on the ratio $|S|/\mu_1$. $\hat V$ is monotonically increasing, so (\ref{eq:VsoftS}) gives the desired runaway. This potential can be calculated analytically for $S$ near the origin ($|S|/\mu_1 \ll 1$), or for large values of the field. Both regimes are interesting for our purpose and we now study them in turn. For a summary on the computation of the two-loop Coleman-Weinberg potential, the reader can refer to appendix \ref{app:VCW}. The results for $S$ very small or large can be obtained by expanding the general formulae in the appendix.

Near the origin the two-loop potential can be expanded in powers of $|S|/\mu_1$, giving~\cite{Giveon:2008wp},
\be\label{eq:Vsoftquad}
V_{soft}\approx -  N\,\left( \frac{h^2}{16 \pi^2}\right)^2\,(h \mu_2^2)^2\,\frac{|S|^2}{\mu_1^2}+ \mc O (|S|^4)\,.
\ee
Therefore $|S|$ acquires a tachyonic mass proportional to $(h \mu_2^2)/\mu_1$ times a two-loop suppression factor. The pseudo-moduli $(X, L, \t L)$ have a positive (one-loop) mass squared of order $m_{CW}^2$ in (\ref{eq:CW}).

For $|S|/\mu_1 \gg 1$, the CW potential exhibits a logarithmic behavior
\be\label{eq:Vsoftlog}
V_{soft}(|S|)\approx- N\,\left( \frac{h^2}{16 \pi^2}\right)^2\,(h \mu_2^2)^2\,\left(\log \frac{|S|^2}{\mu_1^2} \right)^2\,.
\ee
This regime corresponds to the limit of small supersymmetry breaking and the potential can also be obtained from wavefunction renormalization, as follows. For large values of $S$, the fields $\rho_0$ and $\t \rho_0$ in (\ref{eq:Wsinglet}) are the heaviest and can be integrated out supersymmetrically, giving~\cite{Intriligator:2008fe}
\be
K_{eff}= Z_X\,\tr(X^\dag X) + \ldots\;\;,\;\;\log\,Z_X \approx \left( \frac{h^2}{16 \pi^2}\right)^2\,\left(\log \frac{|S|^2}{\mu_1^2} \right)^2\,.
\ee
In the presence of a nontrivial kinetic term metric $G_{X X^\dag}= \partial_X \partial_{X^\dag} K$, the potential is modified to
\be
V \supset G^{X X^\dag} \,W_{X^\dag} W_X = Z_X^{-1}\, W_{X^\dag} W_X\,.
\ee
Then replacing $W_{X_{ij}}= - h \mu_2^2\,\delta_{ij}$ leads to (\ref{eq:Vsoftlog}). This logarithmic regime was discussed in~\cite{Giveon:2008ne} as a way of breaking R-symmetry in metastable vacua. 

In the case $t\neq0$, the dependence of the Coleman-Weinberg potential on $S$ is more complicated, and $S$-dependent terms are generated already at one-loop. However, these are suppressed with respect to the two-loop contributions, due to the symmetry enhancement at $t=0$, for which the one-loop $S$-dependence disappears. The two loop terms still generate a runaway in a similar fashion to the one explained above. (See \S \ref{app:VCW} for more details).

\subsection{Stabilization mechanisms}\label{subsec:stabilize}

Let us next analyze the stabilization of the runaway (\ref{eq:VsoftS}) in the regimes of small and large $|S|/\mu_1$. A local minimum is obtained by turning on a small superpotential perturbation $W \supset t S^k$ and we focus on the renormalizable cases $k=2,\,3$.

First, in the regime of small $S$ a minimum away from the origin ensues when turning on a cubic perturbation. The potential from F-term and soft contributions, ignoring the Higgs fields, becomes
\be
V(|S|) \approx t^2 |S|^4- m_S^2\,|S|^2\;\;,\;\;m_S^2 \approx N\,\left( \frac{h^2}{16 \pi^2}\right)^2\,\left(\frac{h \mu_2^2}{\mu_1}\right)^2\,,
\ee
the soft mass $m_S^2$ following from (\ref{eq:Vsoftquad}) above (when the $t$-dependence of the 2 loop potential is neglected). The combination of these two effects results in a minimum at
\be
|S_0|= \frac{m_S}{\sqrt 2\,t} \approx \sqrt{\frac{N}{2}} \,\frac{h^2}{16 \pi^2}\,\frac{h \mu_2^2}{t\mu_1}\,.
\ee
This gives a consistent solution to the full two-loop potential as long as
\be
t \gtrsim \frac{h^2}{16 \pi^2} \frac{h \mu_2^2}{\mu_1^2}\,,
\ee
which ensures that the minimum $S_0$ is close enough to the origin. For smaller values of the dimensionless coupling $t$, higher order corrections to the CW potential cannot be neglected.

Now we use the general results of \S \ref{subsec:natural} to compute the Higgs parameters. Requiring $B_\mu \sim \mu^2$ sets $\lambda \sim t$ and then
\be
B_\mu \approx \mu^2 \approx\, \frac{N}{2}\,\left( \frac{h^2}{16 \pi^2}\right)^2\,\left(\frac{h \mu_2^2}{\mu_1}\right)^2\,.
\ee
Furthermore, in order to get soft parameters around 1 TeV (and setting for simplicity $\mu_1 \sim \mu_2$), the messenger masses and supersymmetry breaking scale should be
\be
\mu_1 \sim \mu_2 \sim 100 - 200\;{\rm TeV}\,.
\ee
In this range, the gauge-mediated sfermion masses (\ref{eq:mGM}) are naturally of order 1 TeV~\cite{Essig:2008kz}. So a consistent solution to the EWSB conditions ensues. Finally, notice that the dependence on $\lambda$ and $t$ cancels from these expressions, due to the requirement $t \sim \lambda$. However, the physical fluctuations around the vacuum do depend on $\lambda$, and this parameter cannot be too large in order to have a light Higgs above the experimental bound.

Next we focus on the logarithmic regime, and stabilize the runaway using a quadratic $k=2$ superpotential perturbation. (A cubic perturbation can also be considered, but the quadratic case makes the full theory of \S \ref{sec:SQCD} more natural for model-building.) In this case the potential becomes, neglecting again the $t$-dependence of the 2 loop contributions,
\be
V(|S|) \approx t^2 |S|^2- N\,\left( \frac{h^2}{16 \pi^2}\right)^2\,(h \mu_2^2)^2\,\left(\log \frac{|S|^2}{\mu_1^2} \right)^2
\ee
and the minimum is determined iteratively from the condition
\be
|S_0| \approx \sqrt{N}\,\frac{h^2}{8 \pi^2}\,\frac{h \mu_2^2}{t}\,\left(\log \frac{|S_0|}{\mu_1}\right)^{1/2}\,\,.
\ee
The logarithmic description is a good approximation for
\be\label{eq:log-cond}
t \lesssim \sqrt{N}\,\frac{h^2}{8 \pi^2}\,\frac{h \mu_2^2}{\mu_1}\,.
\ee

On the other hand, the conditions (\ref{eq:natural2}) and (\ref{eq:natural}) now translate to
\be
t \sim N^{1/4}\,\frac{h}{4\pi}\,\sqrt{\lambda h \mu_2^2}\,.
\ee
Both constraints are satisfied simultaneously when $\lambda$ is of order of the loop-suppression factor
\be
\frac{\lambda}{h} \lesssim \sqrt{N} \frac{h^2}{8 \pi^2}\,,
\ee
again assuming that $\mu_1 \sim \mu_2$. This small ratio will be explained dynamically in terms of compositeness.

Having satisfied these relations, the Higgs parameters read
\be
B_\mu \sim \mu^2 \sim \sqrt{2N}\,\lambda\frac{h^2}{8 \pi^2}\,h \mu_2^2\,.
\ee
This is of the same order of magnitude as the two-loop gauge mediated contribution (\ref{eq:mGM}). We conclude that the case $k=2$ with a soft logarithmic runaway can naturally give rise to realistic EWSB parameters.

\section{Realization in SQCD with massless and massive flavors}\label{sec:SQCD}

As explained in the Introduction, our aim is to obtain the NMSSM dynamically by identifying the singlet with a confined meson, and simultaneously use this dynamics to obtain a realistic model of gauge mediation. In this section we accomplish this in terms of SQCD in the free magnetic phase, with massless and massive flavors. This will provide a dynamical realization of the O'Raifeartaigh model of direct mediation of \S \ref{sec:OR}, and solve $\mu/B_\mu$ thanks to the mechanism of \S\ref{sec:sol}, where the presence of the runaway is due to the existence of massless flavours in the electric theory \cite{Franco:2006es,Giveon:2008wp}.

The microscopic theory is $SU(N_c)$ SQCD with $N_f$ fundamental flavors $(Q_i, \t Q_i)$. All the electric quarks except for one flavor are given masses $m_i$,
\be
W_{el} = \sum_{i=1}^{N_f-1}\,m_i\, Q_i \t Q_i\,.
\ee
The masses are much smaller than the dynamical scale $\Lambda$, and the massless quarks are $(Q_{N_f}, \t Q_{N_f})$. We require that the massive flavors satisfy
$N_c+2 \le N_f < \frac{3}{2} N_c$,
a condition that is slightly stronger than the free magnetic range restriction. This will be important for the supersymmetry breaking mechanism to apply.\footnote{SQCD with massive and massless quarks, in the context of supersymmetry breaking, was analyzed in refs.~\cite{Franco:2006es}~and~\cite{Giveon:2008wp}. Other applications may be found in~\cite{Giveon:2008ne}. Also, for the mechanism to apply it is not strictly necessary to be in the exact massless limit; rather, it is sufficient if there is a hierarchy of electric masses of order of a loop factor. For simplicity, the exposition is restricted to the massless case, and we note that a small mass term would amount to an additional one-loop contribution to (\ref{eq:VsoftS}).}

The long-distance dynamics admits a Seiberg-dual description~\cite{Seiberg:1994pq} in terms of an 
$SU(\t N_c \equiv N_f - N_c)$  SQCD theory with $N_f$ magnetic dual quarks $(q_i, \t q_i)$ and $N_f^2$ singlets $\Phi$ (corresponding to the rescaled electric mesons $Q_i \t Q_j/\Lambda$), with superpotential
\be
W_{mag}= - h \tr (\hat \mu^2 \Phi) + h \tr (q \Phi \t q)\,.
\ee
The fields have canonical kinetic terms and the relation between electric masses and the matrix $\hat \mu^2$ in the linear term is
\be\label{eq:hatmu}
-h \hat \mu^2 \sim \Lambda\, {\rm diag}(m_1, \ldots, m_{N_f-1},0)\,.
\ee
To reproduce the O'Raifeartaigh model, it will be enough to consider just two different linear terms,
\be\label{eq:hatmu2}
\mu^2 = \left(
\begin{matrix}
\mu_1^2\,\mathbf 1_{\t N_c} & 0 & 0 \\
0 & \mu_2^2 \mathbf 1_{N_f - \t N_c -1} &0\\
0 & 0 & 0 \times \mathbf 1_1
\end{matrix}
\right)
\ee
Here $\mu_1^2> \mu_2^2 > 0$ and $h$ is also chosen to be real.

The superpotential receives nonperturbative contributions controlled by the dynamical scale, creating supersymmetric vacua at
$$
\Phi^{N_f- \t N_c} \sim \mu_i^{2 \t N_c}\,\Lambda^{N_f-3 \t N_c}\,.
$$
On the other hand, the supersymmetry breaking dynamics occurs in the region near the origin $|\Phi| \ll |\Lambda|$, where nonperturbative effects are negligible and the longevity of the vacuum is assured.

\subsection{Metastable supersymmetry breaking}\label{subsec:vac}

Near the origin supersymmetry is broken,
\be
W_{\Phi^T}= - h \hat \mu^2 + h \,q \t q \neq 0
\ee
because the second term has smaller rank than the first. The stable minimum corresponds to cancelling the largest entries in the matrix $\hat \mu^2$ with expectation values of $q \t q$,
\be
\langle q \t q \rangle = \mu_1^2\,\mathbf 1_{\t N_c}\;\;,\;\;\langle W_{\Phi} \rangle = - h \mu_2^2\,\mathbf 1_{N_f - \t N_c -1}\,.
\ee
Taking into account nonperturbative effects, this vacuum becomes metastable and long-lived as long as $|m_i| \ll |\Lambda|$.

We see that, while the first $\t N_c$ flavors of magnetic quarks acquire a VEV and are approximately supersymmetric, there are $N_f - \t N_c -1$ flavors that couple directly to the nonzero F-terms, plus a single flavor $(q_{N_f}, \t q_{N_f})$ without tree-level couplings to the F-terms. This is precisely the structure that we need for our mechanism, provided that the singlet $S$ is identified with the $(N_f, N_f)$ element of the meson
$$
S = \frac{Q_{N_f} \t Q_{N_f}}{\Lambda}\,.
$$

In more detail, the fluctuations around the vacuum are parametrized by
\be\label{eq:fluctPhi}
\Phi = \left( 
\begin{matrix}
Y_{\t N_c \times \t N_c} & Z^T_{\t N_c \times (N_f - \t N_c -1)} & (Z_0^T)_{\t N_c \times 1} \\
\t Z_{(N_f - \t N_c -1) \times \t N_c} & X_{(N_f - \t N_c -1)  \times (N_f - \t N_c -1) } & L_{(N_f - \t N_c -1)  \times 1} \\
(\t Z_0)_{1 \times \t N_c} & \t L_{1 \times (N_f - \t N_c -1) } & S_{1 \times 1}
\end{matrix}
\right)\;\;,
\ee
\be
\t q =\left(
\begin{matrix}
\t \chi_{\t N_c \times \t N_c} \\ \t \rho_{(N_f - \t N_c -1)  \times \t N_c} \\ (\t \rho_0)_{1 \times \t N_c}
\end{matrix}\right)\;\;,\;\;
q =\left(
\begin{matrix}
\chi _{\t N_c \times \t N_c}\\ \rho_{(N_f - \t N_c -1)  \times \t N_c} \\ (\rho_0)_{1 \times \t N_c}
\end{matrix}
\right)
\ee
where the subindices indicate the sizes of the corresponding matrices.

The fields $X,\, L,\, \t L$ and $S$ are pseudo-moduli; all the other fields are stabilized at the origin, except for $\langle \chi \t \chi \rangle = \mu_1^2 \, \mathbf 1_{\t N_c}$. $(Y, \chi, \t \chi)$ are supersymmetric at tree level and can be integrated out, leaving
\begin{eqnarray}
W & = & - h \mu_2^2\,\tr\,X + h\,\tr(\rho X \t \rho) + h\,\rho_0 S \t \rho_0 + h \,\tr(\rho L \t \rho_0+ \rho_0 \t L \t \rho)+ \nonumber\\
&+& h \mu_1\,\tr(\rho \t Z + \t \rho Z)+ h \mu_1\,(\rho_0 \t Z_0 + \t \rho_0 Z_0)\,.
\end{eqnarray}
For $\t N_c = 1$ and $N_f - \t N_c -1 = N$, this is precisely the superpotential  (\ref{eq:Wtotal}) for the O'Raifeartaigh model that gives the desired runaway. The $\mc O(h \mu_1)$ supersymmetric masses come from the VEVs of $\chi$ and $\t \chi$, and the cubic coupling between the two types of messengers $(\rho, Z)$, $(\rho_0, Z_0)$ and link fields arise from the term $q \Phi \t q$ dictated by Seiberg duality.

\subsection{Deformations and R-symmetry breaking}\label{subsec:Rbreaking}

Finally, let us describe the deformations of the electric theory that lead to Eq.~(\ref{eq:Weff}). These appear from higher-dimensional operators in the electric theory generated at a certain scale $\Lambda_0 > \Lambda$,
$$
\Delta W_{el} = \frac{\alpha_1}{\Lambda_0} (Q_{N_f} \t Q_{N_f}) H_u H_d +\frac{\alpha_2}{\Lambda_0^{2k-3}}\,(Q_{N_f} \t Q_{N_f})^k\;\Rightarrow\;
\Delta W_{mag}= \lambda S H_u H_d + \frac{t}{k} S^k\,.
$$
with $\lambda \approx \alpha_1 \Lambda/\Lambda_0$ and $t \approx \alpha_2 \Lambda^k/\Lambda_0^{2k-3}$.

The case $k=3$ where $S$ appears with a cubic coupling is unnatural in this framework: a realistic Higgs phenomenology requires $\lambda \sim t$, but these correspond to dimension 4 and 6 operators in the electric theory, respectively. It may be possible to generate this structure using nonperturbative superpotentials, but we will not explore this further here.

Let us then focus on the case $k=2$. This situation is quite interesting, because a perturbation $W_{mag} \supset \tr \Phi^2$ in the magnetic theory would simultaneously produce realistic Higgs parameters and gaugino masses from the breaking of the R-symmetry. This was studied in detail in~\cite{Essig:2008kz}. The full electric superpotential is\footnote{The dimensionless parameter $\alpha_2$ needs to be a small number; this is natural from the point of view of the symmetries of the theory. Also, a double-trace perturbation $(\tr\, Q \t Q)^2$ is required if the SM generations are elementary~\cite{Essig:2008kz}.}
\be
W_{el}=\tr(m Q \t Q) + \frac{\alpha_1}{\Lambda_0}\,(Q \t Q)_{N_f} H_u H_d+ \frac{\alpha_2}{2\Lambda_0}\,\tr(Q \t Q)^2\,.
\ee
Below the dynamical scale $\Lambda$ the theory confines and the canonically normalized meson corresponds to
\be
\Phi = \frac{Q \t Q}{\Lambda}\,.
\ee
The superpotential of the magnetic theory becomes
\be\label{eq:Wmag3}
W_{mag}=- h \,\tr (\hat \mu^2 \Phi) + h \,\tr (q \Phi \t q) + \frac{1}{2} h^2 \muphi \,\tr \,\Phi^2 + \lambda \Phi_{N_f,N_f} H_u H_d
\ee
The linear term is related to the masses by (\ref{eq:hatmu}), and the particular choice (\ref{eq:hatmu2}) is taken. On the other hand, $\muphi \approx \alpha_2 \Lambda^2/\Lambda_0$. In the long-distance theory, $\lambda \ll 1$ and $\muphi \ll \mu$ naturally, because they appear as irrelevant perturbations to the UV dual.

The quadratic term breaks explicitly the R-symmetry and, using the parametrization of Eq.~(\ref{eq:fluctPhi}), it gives rise to the desired $X^2+S^2$ interactions. This perturbation introduces supersymmetric vacua at
\be
h \langle \Phi_{susy} \rangle = \frac{\hat \mu^2}{\muphi}\,.
\ee
As long as the perturbation is small enough,
\be\label{eq:meta-constraint}
\muphi^2 \lesssim \frac{1}{16 \pi^2} \frac{\mu_2^4}{\mu_1^2}\,,
\ee
there is still a long-lived metastable vacuum, albeit displaced from the origin by
\be
 h X_0 \approx 16 \pi^2\, \frac{\mu_1^2}{\mu_2^2}\, \muphi\,.
\ee
The spontaneous breaking $h X_0$ of the $U(1)_R$ is a loop factor larger than the explicit breaking parameter $\muphi$.

The stabilization of $S$ was described in \S \ref{subsec:stabilize}; the parameter $t$ there corresponds to $t=h^2\muphi$. The quadratic supersymmetric potential is stabilized against the logarithmic runaway from the two-loop potential --whose computation is summarized in \S \ref{app:VCW}-- yielding
\be\label{eq:S0-muphi}
|S_0| \approx \sqrt{2(N_f-\t N_c-1)}\,\frac{h^2}{8 \pi^2}\,\frac{\mu_2^2}{h \muphi}\,.
\ee
The condition (\ref{eq:log-cond}) for the logarithmic approximation to be valid is
\be\label{eq:log-constraint}
h^2 \muphi \lesssim \sqrt{N_f-\t N_c-1}\,\frac{h^2}{8 \pi^2} \,\frac{h \mu_2^2}{\mu_1}\,.
\ee
This is compatible with the metastability constraint (\ref{eq:meta-constraint}), and in fact it is stronger if $h \lesssim 2\pi / \sqrt{2(N_f-\t N_c-1)}$.

\subsection{Microscopic corrections}\label{subsec:microscopic}

Having analyzed one- and two-loop effects in the macroscopic theory, we need to make sure that our conclusions are not affected by corrections from the microscopic theory. These appear from loops of short-distance modes at the dynamical scale $\Lambda$. In particular, the stabilization of $S$ depends crucially on small two-loop effects; these could be sensitive to microscopic corrections.

Integrating out heavy modes at the scale $\Lambda$ produces corrections to the K\"ahler potential of the form
\be\label{eq:delK}
\delta K = c \frac{|\Phi|^4}{|\Lambda|^2} + \ldots
\ee
where $c$ is an incalculable dimensionless constant.\footnote{The effect of (\ref{eq:delK}) on $X$ was already studied in~\cite{Intriligator:2006dd}.} This changes the kinetic term metric, correcting the potential by
\be\label{eq:delV}
\delta V = -c \frac{|\Phi|^2}{|\Lambda|^2}\,(h \mu_2^2)^2 +\ldots
\ee

First, microscopic corrections to the stabilization of $X$ can be neglected if (\ref{eq:delV}) is much smaller than $V \supset m_{CW}^2 |X|^2$, namely
\be
\frac{\mu_1^2}{\Lambda^2} \ll \frac{1}{16 \pi^2} \,.
\ee
On the other hand, microscopic corrections to $S$ may be ignored if they are smaller than $V \supset |h^2 \muphi|^2 |S|^2$,
\be\label{eq:cond-Lambda}
\frac{\mu_2^2}{\Lambda} \ll \muphi\,.
\ee
As long as these conditions are met, microscopic corrections to the metastable structure are negligible. These will be satisfied in the concrete model below.

\section{Phenomenology of the $k=2$ model}\label{sec:k2}

In this last section we describe the low energy phenomenology of the $k=2$ model, with superpotential (\ref{eq:Wmag3}), where $\Phi$ is given in (\ref{eq:fluctPhi}), $\Phi_{N_f,N_f}\equiv S$, and $\tr \,\Phi^2 \supset Y^2 + \tr \,X^2 + S^2$. In this model the NMSSM singlet appears with a superpotential mass term, which is naturally small because it comes from an irrelevant perturbation in the dual electric theory. Furthermore, this mass scale also triggers R-symmetry breaking and generates nonzero gaugino masses. Since the breaking of R-symmetry is through a small parameter, the dangerous K\"ahler potential operator of Eq.~\eqref{eq:Stadpoles} will be naturally suppressed.

The minimal case corresponds to an electric theory with $N_c=6$ colors and $N_f=7$ flavors. The electric masses break $SU(N_f=7)$ to $SU(5)$ acting on $X$ and the messengers $(\rho, Z, L)$. This $SU(5)$ global symmetry (or an $SU(3) \times SU(2) \times U(1)$ subgroup thereof) is weakly gauged and identified with the SM gauge group. The baryon number is also gauged to remove a NG boson. Notice that the magnetic gauge group is trivial.

\subsection{Spectrum}\label{subsec:spectrum}

Let us present the spectrum in terms of the superpotential parameters, starting from the heavier fields. For reference, the supersymmetry breaking fluctuations around the metastable vacuum were given in \S \ref{subsec:vac}, while the Higgs sector masses and interactions with the singlet were presented in \S \ref{subsec:singlet}.\footnote{The supersymmetry breaking spectrum was studied in detail in~\cite{Essig:2008kz}, so our analysis of some of the fluctuations will be brief.} The masses are given at the messenger scale.

\vskip 1mm

\noindent $\bullet$ First, the fields $Y$ and $\chi+ \t \chi$ have tree-level masses of order $\sim h \mu_1$. ${\rm Re}(\chi - \t \chi)$ is a pseudo-modulus stabilized at one-loop, so it acquires a mass $m_{CW}$ as in (\ref{eq:CW}). ${\rm Im}(\chi - \t \chi)$ is the NG boson which gets a mass $g_V \mu_1$ by gauging $U(1)_V$.

\vskip 1.5mm

\noindent $\bullet$ The fields responsible for transmitting the breaking of supersymmetry come from the $(\rho, Z)$ sector. They have supersymmetric masses $\sim h \mu_1$ with soft splittings of order $h \mu_2$. There are also NG bosons, which become massive after weakly gauging the $SU(5)$ symmetry. On the other hand, the fields $(\rho_0, Z_0)$ are approximately supersymmetric, with masses $\sim h \mu_1$.

\vskip 1.5mm

\noindent $\bullet$ Moving now to the remaining light fields in the meson $\Phi$, the real part of the pseudo-modulus $X$ has a one-loop mass squared $m_{CW}^2$, as we explained in (\ref{eq:CW}). The ``link'' fields $(L, \t L)$ also acquire one-loop masses, but their coupling to the messengers are suppressed by $\mu_1/S_0$ (this can be seen by integrating out $(\rho_0, Z_0)$ supersymmetrically in $W_{mag}$). Thus their masses are of order 
$$
m^2 \sim \frac{(h \mu_1)^2}{|S_0|^2} m_{CW}^2\,.
$$ 
For the realistic choice of parameters below, we will have $ |S_0| \sim 10 \times h \mu_1$, implying that $(L, \t L)$ have masses one order of magnitude lighter than the CW scale. The meson component corresponding to the singlet $S$ is the lightest. Its real part has a mass squared $m^2 \sim (h^2 \muphi)^2$; this corresponds to a two-loop effect, which can be seen by relating $\muphi$ to $\mu_1$ and $\mu_2$ by (\ref{eq:log-constraint}). 

The R-symmetry is explicitly broken by $\muphi$, giving masses to the phases of $X$ and $S$. The normalized axion from $X$ corresponds to $a_X \sim {\rm arg}(X)/|X_0|$. It acquires a mass
\be
\label{eq:Xaxion}
m_{a_X}^2 = 2 \frac{  h \mu_2^2 (h^2 \muphi)}{\sqrt{N_c-1} X_0} \sim \frac{h^2}{16 \pi^2} \frac{(h \mu_2^2)^2}{\mu_1^2}\,,
\ee
which is comparable with the Coleman-Weinberg scale for $\mu_1 \sim \mu_2$.

On the other hand, the dominant mass contribution to $a_S \sim {\rm arg}(S)/|S_0|$ comes from the two-loop potential, as discussed on appendix \ref{app:VCW}. For the appropriate window of values for $|S_0|$ and $h\sim \mathcal O(1)$ it is of the order of
\be\label{eq:maS}
m_{a_S}^2 \sim\frac{\mu_2\mu_\phi}{(16\pi^2)^2}.
\ee

\vskip 1.5mm

\noindent $\bullet$ Lastly, the MSSM gauginos and sfermions acquire one- and two-loop masses, respectively, from the usual gauge mediated diagrams. For gauginos we find
\be
m_\lambda \sim g_{SM}^2 \,\frac{\mu_2^4}{\mu_1^4}\,\muphi
\ee
where the one-loop factor cancelled the loop factor in $X_0 \sim 16 \pi^2 \muphi$ --there are also contributions that depend on $S_0$, but they are subleading. For sfermions,
\be
m_{GM}^2 \sim \left(\frac{g_{SM}^2}{16 \pi^2}\right)^2\,\frac{(h \mu_2^2)^2}{\mu_1^2}\,.
\ee
These masses are comparable if
\be\label{eq:muphi-pheno}
\muphi \sim \frac{1}{16 \pi^2} \frac{h \mu_1^3}{\mu_2^2}
\ee
which is marginally compatible with the condition (\ref{eq:log-constraint}).

The soft Higgs masses $m_{H_u}^2$ and $m_{H_d}^2$ are given by $m_{GM}^2$ at the messenger scale, while $\mu$ and $B_\mu$ become
\be
\mu = \sqrt{2(N_f-2)} \frac{h^2}{8 \pi^2} \frac{\lambda \mu_2^2}{h \muphi}\;\;,\;\;B_\mu=-\sqrt{2(N_f-2)} \frac{h^2}{8 \pi^2}\,\lambda h \mu_2^2\,.
\ee
Given (\ref{eq:muphi-pheno}), $B_\mu \sim \mu^2$ for
\be
\lambda \approx \frac{1}{\sqrt{2(N_f-2)}} \frac{h^2}{8 \pi^2} \frac{h \mu_1^6}{\mu_2^6}\,.
\ee
For a supersymmetry breaking scale around $100 -200$ TeV, $m_{H_u}^2$ is driven tachyonic due to RG and finite corrections, and the EW symmetry is broken radiatively. 

\subsection{Realistic parameter choices}\label{subsec:parameters}

Let us set $\mu_1 \sim \mu_2$. Gaugino masses at or below 1 TeV are obtained if 
\be
\muphi \sim 1\,{\rm TeV}\,;
\ee
having comparable sfermion masses (see Eq.~(\ref{eq:muphi-pheno})) then fixes the scale of supersymmetry breaking at
\be
\sqrt{F} \sim \mu_2 \sim 100 - 200\;{\rm TeV}\,.
\ee
In numerical examples, the sfermions are typically heavier than the gauginos by a factor of 2 or 3~\cite{Craig:2009hf,SchaferNameki:2010iz}.

This class of models gives low scale supersymmetry breaking. The corresponding gravitino mass is found to be
\be
m_{3/2} \approx \frac{F}{\sqrt 3\,M_{P}} \sim 1 - 10\;{\rm eV}
\ee
where $M_{P} \approx 2.4 \times 10^{18}\;{\rm GeV}$ is the reduced Planck mass. This light gravitino is in agreement with cosmological constraints. 

Above the TeV scale there is additional matter charged under the SM gauge group. For these choices of parameters, the Coleman-Weinberg scale is
\be
m_{CW} \sim 10 - 15\;{\rm TeV}\,.
\ee
This is the mass for the $SU(5)$ adjoint (plus singlet) $X$. Furthermore the messenger fields $(\rho, Z, \t \rho, \t Z)$ have masses around $\sim 100 - 200\;{\rm TeV}$. By minimizing the full two-loop Coleman Weinberg potential of appendix \ref{app:VCW} (including the $\muphi$ interactions), the expectation value of $S$ is found to be around
\be
S \sim\;(1 -3)\;\times 10^3\;{\rm TeV}\,, 
\ee
in parametric agreement with the analytic analysis of \S \ref{subsec:Rbreaking}. In order to get acceptable values for $\mu$ and $B_\mu$ at the TeV scale, one should choose $\lambda\sim10^{-3}$. The model in general predicts small $\tan\,\beta$, without large hierarchies in the Higgs sector.

The spectrum of the model presents two additional interesting features. First, the next lightest particle above the gravitino corresponds to the axion (\ref{eq:maS}), with mass around $50 - 100\;{\rm GeV}$. This neutral particle has interactions with the messengers $(\rho_0,Z_0)$, which may provide an interesting `portal' into the hidden sector \cite{Cheung:2010jx}. Furthermore, the link fields $(L, \t L)$, which are fundamentals under the SM gauge group, have masses of the order of 1 TeV. At the scale of the light MSSM sfermions, we then also find a `vector-like' generation. These fields are naturally light because of the R-symmetry and suppression by the large expectation value of $S$. It would be interesting to understand the phenomenological consequences of these two aspects.

Finally, the condition (\ref{eq:cond-Lambda}) for microscopic corrections to be negligible requires the dynamical scale $\Lambda$ to be larger than $\mu_2^2/\muphi \sim 10^4 $ TeV. On the other hand, given the above matter content, there is a Landau pole for $SU(3)_C$ at around $10^6$ TeV and, for consistency of the microscopic theory, $\Lambda$ needs to be below the Landau pole. Both requirements are then nontrivial, combining to define an allowed range of 
\be
10^4\;{\rm TeV} < \Lambda < 10^6\;{\rm TeV}\,.
\ee 
The main contributions to the Landau pole come from the SM matter in $X$. As pointed out in~\cite{Franco:2009wf}, the Landau pole can be pushed near the GUT scale by introducing additional singlets $S_{\bf R}$ with cubic couplings $W \supset  (Q \t Q)_{\overline {\bf R}} S_{\bf R}$, which gives masses of order $ \Lambda$ to unwanted matter from $X$. This introduces new metastable vacua, but our vacuum is energetically preferred if $\mu_2 < \mu_1$~\cite{Behbahani:2010wh}. In this case, $\Lambda$ can be chosen much larger as well --albeit at the cost of introducing by hand extra matter $S_{\bf R}$.

\subsection*{Acknowledgements}

We would like to thank 
M.~Dine,
R.~Essig,
S.~Kachru,  and
J.~Wacker
for very interesting discussions and comments. S.S.-N. and C.T. are supported in part by NSF grant PHY-05-51164 at the KITP.
G.T. is supported by the US DOE under contract number DE-AC02-76SF00515 at SLAC.
C.T. is supported by  in part by MICINN through grant
FPA2008-04906 and by both MICINN and the Fulbright Program through grant 2008-0800.
G.T. and C.T. thank the KITP,  and S.S.-N. thanks the Caltech Theory group, for their generous hospitality.

\appendix

\section{Estimation of the mass shift of the light Higgs}\label{app:mass-shiftSSN}

The mass eigenvalues can be studied treating the off-diagonal matrix elements mixing the singlet with the Higgs fields as perturbations. To this end, we parameterize the neutral scalar mass matrix of Eq.~\eqref{eq:masses} as
\be
M_S^2 = \left( 
\begin{array}{ccc}
m^2_{11} &     m^2_{12} & \epsilon_1 \\
m^2_{12} &    m^2_{22} & \epsilon_2 \\
\epsilon_1 & \epsilon_2 &   m_S^2
\end{array}
\right)\,,
\ee 
where the $\epsilon_i$ are treated as perturbations. To estimate the size of the matrix elements, we impose naturalness and use Eq.~\eqref{eq:masses}; it is easily seen that in this case the elements $m^2_{ij}$ are also of order $\mu^2$. Thus, for $\epsilon_i=0$ one obtains mass eigenvalues for the neutral Higgs fields $h$ and $H$ of the same order,
\be
m_H^2 \sim m_h^2 \sim  O(\mu^2) \,.
\ee

To compare with the MSSM result, it can be seen that, for $B_\mu\sim\mu^2$ 
\be
m_h^2=m_{h,MSSM}^2+\delta_1 m_h^2+\delta_2 m_h^2\,,\quad\qquad\qquad  \delta_1 m^2_h\sim\lambda^2 v^2 \sin^2(2\beta)\,,
\ee
and $\delta_2 m_h^2$ is produced by the $\epsilon_i$ perturbations, 
\be
\delta_2 m^2_h = -\frac{\epsilon _1^2 \left(m_h^2-m^2_{22}\right){}^2+\epsilon_2 ^2 m^4_{12}+2\epsilon_1\epsilon_2 
\left(m_h^2-m^2_{22}\right)m^2_{12}}{\left(m_{S}^2 -m_h^2\right)\left(\left(m_h^2- m^2_{22}\right)^2+m_{12}^4\right) }\,.
\ee
The numerator of this expression is strictly positive, and so will be the denominator for natural values of $\mu>m_h$ of the order of the electroweak scale. Thus $\delta_2 m^2$ is strictly negative for phenomenologically acceptable vacuum configurations. Neglecting $m^2_h$ against the other contributions and setting $B_\mu = B \mu^2$ we obtain
\be
\begin{aligned}
m^2_{22}& \sim  B \mu^2 \tan\beta\cr
m^2_{12}& \sim - B \mu^2 \cr
\epsilon_1&\sim  v \lambda \mu (2 \sin\beta - (k-1) B \cos\beta)  \cr
\epsilon_2&\sim  v \lambda \mu (2 \cos\beta -(k-1) B \sin\beta )  \,.
\end{aligned}
\ee
The denominator of $\delta_2m^2_h$ in this approximation takes the form
\be
m_S^2 (m_{22}^4 + m_{12}^4) \sim 2 ((k-1)(k-2)B^2 +1) B \mu^4  (\tan^2\beta +1)
\ee
and the numerator simplifies to
\be
\epsilon_1^2 m_{22}^4 + \epsilon_2^2 m_{12}^4 - 2 \epsilon_1 \epsilon_2 m_{22}^2 m_{12}^2 
\sim 
v^2 \lambda^2B \mu^4 (\sin^2\beta  (\tan\beta +1)^2 ((k-1)B -2)^2) \,.
\ee

Putting it all together we obtain
\be
\delta_2 m_h^2 
\sim - \frac{v^2 \lambda^2}{2}   \frac{(B (k-1)\sin(2 \beta)- 2)^2}{ (k-1 )(k-2) B^2 +1} \,.
\ee
i.e.,
\be
(\delta_1+\delta_2) m_h^2  \sim - v^2 \lambda^2 f_k(\beta) \,  \,,
\ee
where $f_k(\beta)$ is an order one function,
\be
f_k(\beta)\sim  \frac{1}{2}   \frac{(B (k-1)\sin(2 \beta)- 2)^2}{ (k-1 )(k-2) B^2 +1} -\sin^2(2\beta)  \,.
\ee
The sign of the shift depends on the $B$; 
for example for $k=2$, $B=3$, $\tan\beta>4$ the shift is negative. These effects are negligible in the model of \S \ref{sec:k2}, for which $\lambda \sim 10^{-3}$, but they could give important effects in the case $k=3$.

\section{Coleman-Weinberg potential}\label{app:VCW}

The vacuum structure of the theory and the masses of the pseudo-moduli fields $X$ and $S$, follow from the Coleman-Weinberg potential of the fields that couple to them. In this appendix we summarize this calculation for the SQCD model of \S \ref{sec:SQCD} in the $\rm\overline{MS}$ scheme, using the conventions in ref.~\cite{Martin:2001vx}. 

The relevant superpotential terms in the low-energy magnetic theory are those given in Eq.~\eqref{eq:Wmag3}, where the fluctuations of the fields $\Phi,q,\tilde q$ are decomposed as in Eq.~\eqref{eq:fluctPhi}, and in the vacuum one has $\langle \chi \t \chi \rangle = \mu_1^2 \, \mathbf 1_{\t N_c}$. The fields $(Y,\tilde\chi,\chi)$ are supersymmetric at tree-level and do not couple to the moduli; hence they are not relevant for our CW calculations and can be integrated out; this yields \begin{align}\nonumber
W=&-h\mu_2^2\tr X+h\tr(\rho X\tilde \rho) + h\,\rho_0 S \t \rho_0 + h \,\tr(\rho L \t \rho_0+ \rho_0 \t L \t \rho)
+ h \mu_1\,\tr(\rho \t Z + \t \rho Z)\\
&+ h \mu_1\,(\rho_0 \t Z_0 + \t \rho_0 Z_0)+\frac{h^2}{2}\mu_\phi(X^2+S^2+2Z^T\t Z+2Z_0^T\t Z_0+L\t L).
\label{eq:Wtot}
\end{align}

\subsection*{Mass eigenstates}

In order to compute the CW potential generated by the fields $(\rho,\t\rho,Z,\t Z,\rho_0,\t\rho_0,Z_0,\t Z_0,L,\t L)$, we need their moduli-dependent mass matrices. We group all fields except $L,\tilde L$ in  the following field multiplets,
\begin{align*}
 \h{\Phi}=\left[\begin{array}{c}
                 \rho \\ Z \\ \t\rho^* \\\t Z^*
                \end{array}\right],\,\,
 \h{\Phi}_0=\left[\begin{array}{c}
                 \rho_0 \\ Z_0 \\ \t\rho^*_0 \\\t Z^*_0
                \end{array}\right],\,\,
\h{\t\Psi}=\left[\begin{array}{c}
                 \t\Psi_\rho \\ \t\Psi_Z
                \end{array}\right],\,\
\h{\Psi}=\left[\begin{array}{c}
                 \Psi_{\rho} \\ \Psi_{Z}
                \end{array}\right],\,\
\h{\t\Psi}_0=\left[\begin{array}{c}
                 \t\Psi_{\rho_0} \\ \t\Psi_{Z_0}
                \end{array}\right],\,\
\h{\Psi}_0=\left[\begin{array}{c}
                 \Psi_{\rho_0} \\ \Psi_{Z_0}
                \end{array}\right],\,\
\end{align*}
where, for example, $Z$ and $\Psi_Z$ denote, respectively, the scalar and fermionic components of the chiral superfield $Z$, and similarly for the other fields. With this notation, the mass terms for the above multiplets and the scalar and fermionic components of the fields $L,\tilde L$ are
\begin{align*}
 {\cal L}\supset -\hat\Phi^\dagger \h M^2_b\h\Phi-\hat\Phi_0^\dagger \h M^2_{b,0}\h\Phi_0-h^4\muphi^2(\tilde L^\dagger\tilde L+L^\dagger L)-(\hat{\t\Psi} \h M_f\h\Psi+\hat{\t\Psi}_0 \h M_{f,0}\h\Psi_0+h^2\mu_\phi\hat{\t\Psi}_L \h\Psi_L+c.c.),
\end{align*}
\begin{align} 
 \h M_f\!=\!\left[\begin{array}{cc}
            h X & h \mu_1\\ h\mu_1 & h^2\mu_\phi
           \end{array}\right]\!,
\h M_{f0}\!=\!\left[\begin{array}{cc}
            \!h S & h \mu_1\\ \!h\mu_1 & h^2\mu_\phi
           \end{array}\right]\!,
\h M^2_{b}\!=\!\left[\begin{array}{cc}
           M_f^\dagger M_f & -h F^\dagger_X\\ -hF_X& M_f M^\dagger_f
           \end{array}\right]\!,
\h M^2_{b,0}\!=\!\left[\begin{array}{cc}\!
           M_{f,0}^\dagger M_{f,0} & -h F^\dagger_S\\\! -hF_S& M_{f,0} M^\dagger_{f,0}
           \end{array}\right]\!,
\label{eq:massmatrices}
\end{align}
with 
\begin{align*}
 -F^\dagger_X=h\left[\begin{array}{cc}
                      -\mu_2^2+h\mu_\phi X&0\\
		      0 & 0
                     \end{array}\right],\,\,
-F^\dagger_S=h\left[\begin{array}{cc}
                      h\mu_\phi S&0\\
		      0 & 0
                     \end{array}\right].
\end{align*}

The superfields $L$ and $\tilde L$ have a supersymmetric spectrum, with a common mass of $h^2\muphi$. It is useful to rotate the mass matrices in eq.~\eqref{eq:massmatrices} to the eigenvalue basis. The corresponding bosonic mass eigenvalues squared, as well as the fermionic eigenvalues for the matrices $M^\dagger_f M_f$, are, with $\sigma,\eta\in\{1,-1\}$,
\begin{eqnarray}\label{eq:nonSUSYmasses}
m^2_f&&= h^2\mu_1^2+\frac{1}{2}h^2|X|^2+\frac{1}{2}h^4 \muphi^2+\frac{1}{2}\sigma\sqrt{\left(|hX|^2-|h^2 \muphi|^2\right)^2+4|h^2\mu_1 X^*+h^3 \mu_1\muphi|^2}\nonumber\\
m^2_{f,0}&&= h^2\mu_1^2+\frac{1}{2}h^2|S|^2+\frac{1}{2}h^4 \muphi^2+\frac{1}{2}\sigma\sqrt{\left(|hS|^2-|h^2 \muphi|^2\right)^2+4|h^2\mu_1 S^*+h^3 \mu_1\muphi|^2}\nonumber\\
m^2_b&& = h^2\mu_1^2+\frac{1}{2}|hX|^2+\frac{1}{2}h^4\muphi^2+\frac{1}{2}\eta|h^2\mu_2^2-h^3 \muphi X|\nonumber \\
 && +\frac{1}{2}\sigma\left[\left(|hX|^2-|h^2 \muphi|^2+\eta|h^2\mu_2^2-h^3 \muphi X|\right)^2+ 4|h^2\mu_1X^*+h^3\mu_1 \muphi|^2\right]^{1/2}\,\nonumber\\
m^2_{b,0}&& = h^2\mu_1^2+\frac{1}{2}|hS|^2+\frac{1}{2}h^4\muphi^2+\frac{1}{2}\eta h^3\muphi|S|\nonumber \\
 && +\frac{1}{2}\sigma\left[\left(|hS|^2-|h^2 \muphi|^2+\eta h^3 \muphi|S|\right)^2+ 4|h^2\mu_1S^*+h^3\mu_1 \muphi|^2\right]^{1/2}.
\end{eqnarray}

\subsection*{CW potential at two-loops}

The CW computation can be expressed, as in ref.~\cite{Martin:2001vx}, in terms of the couplings of the theory written on a basis of real scalar fields and Weyl fermions in which the bosonic mass matrices, as well as the fermionic mass matrices squared $M^\dagger_f M_f$, are diagonal. We will denote these scalar and fermionic fields as $\phi_i,\psi_I$, respectively. The mass terms and interaction Lagrangian can be written as
\begin{align*}
 {\cal L}\supset-\frac{1}{2}\phi_i M^2_b\phi_i-\frac{1}{2}\psi_I M_{f,IJ}\psi_J-\frac{1}{6}\lambda^{ijk}\phi_i\phi_j\phi_k-\frac{1}{24}\lambda'^{ijkl}\phi_i\phi_j\phi_k\phi_l-\Big(\frac{1}{2}Y^{IJk}\psi_I\psi_J\phi+c.c.\Big),
\end{align*}
where $\lambda^{ijk}$ and $\lambda'^{ijkl}$ are completely symmetric under permutations of their indices, and $Y^{IJk}=Y^{JIk}$, $M_{f,IJ}=M_{f,JI}$. The mass eigenvalues are those of (\ref{eq:nonSUSYmasses}), while the couplings $\lambda, \lambda', Y$ can be obtained from both the superpotential in \eqref{eq:Wtot} as well as the matrices that relate the fields $\h\Phi,\h\Psi$ with $\phi,\psi$. The expressions for arbitrary $\mu_\phi$ are quite lengthy and will not be given here.
  
Writing the loop expansion of the Coleman-Weinberg potential as
\begin{align*}
 V=V^{(0)}+\frac{1}{16\pi^2}V^{(1)}+\frac{1}{(16\pi^2)^2}V^{(2)}+\dots,
\end{align*}
then in the $\overline{\rm MS}$ scheme, denoting the bosonic and fermionic mass square eigenvalues as $m_{b,i}^2,m_{f,i}^2$, respectively, one has \cite{Martin:2001vx}
\begin{align}
\nonumber &V^{(1)}(Q^2)=\frac{1}{4}(m_{b,i}^2)^2\Big(\log\frac{m_{b,i}^2}{Q^2}\Big)-\frac{1}{2}\sum_I(m_{f,I}^2)^2\Big(\log\frac{m_{b,i}^2}{Q^2}\Big),\\
\nonumber &V^{(2)}(Q^2)=V^{(2)}_{SSS}+V^{(2)}_{SS}+V^{(2)}_{FFS}+V^{(2)}_{SSV},\\
\label{eq:VCW2}&V^{(2)}_{SSS}=\frac{1}{12}(\lambda^{ijk})^2 f_{SSS}(m^2_i,m^2_j,m^2_k,Q^2),\\
\nonumber &V^{(2)}_{SS}=\frac{1}{8}\lambda^{iijj} f_{SS}(m^2_i,m^2_j,Q^2),\\
\nonumber &V^{(2)}_{FFS}=\frac{1}{2}|Y^{IJk}|^2 f_{FFS}(m^2_I,m^2_J,m^2_k,Q^2),\\
\nonumber &V^{(2)}_{\overline{FF}S}=\frac{1}{4}Y^{IJk}Y^{I'J'k}M^*_{II'}M^*_{JJ'}f_{\overline{FF}S}(m^2_I,m^2_J,m^2_k,Q^2)+c.c.,
\end{align}
where $Q$ is the renormalization scale, all indices are summed, and the functions $f_{SSS}$, $f_{SS}$, $f_{FFS}$ and $f_{\overline{FF}S}$ can be found on ref.~\cite{Martin:2001vx}.\footnote{A subtlety here is that some of the expressions given there for $f_{a}(x,y,z)$ are only valid for $x^2+y^2+z^2-2xy-2xz-2yz>0$, which can be violated for typical choices of parameters in the masses of \eqref{eq:nonSUSYmasses}. To circumvent this one may use the alternative expressions given for example in~\cite{Ford:1992pn}.}

From the superpotential \eqref{eq:Wtot} and the  expressions in \eqref{eq:massmatrices} for the mass matrices, all S-dependence in the Coleman-Weinberg potential comes from  the fluctuations $\hat\Phi_0$, be it through their masses or their contributions to the couplings $\lambda,\lambda',Y$. In the case $\muphi=0$, the tree-level spectrum of the fields $\hat\Phi_0$ is supersymmetric, so their the one-loop contribution vanishes identically. Thus, all S-dependence will arise at two-loops and beyond, while $X$ is stabilized at one-loop as described in \S \ref{subsec:review}.
The two-loop potential~\cite{Giveon:2008wp} generates the runaway for $S$ mentioned in \S \ref{subsec:runaway}; the behavior of the potential near the origin $S=0$ and for large $S/\mu_1$ follows by appropriately expanding (\ref{eq:VCW2}) and gives~\eqref{eq:Vsoftquad}~and~\eqref{eq:Vsoftlog}, respectively. There are massless axion fields corresponding to the phases of the fields $X$ and $S$, associated to the breaking of chiral symmetries.

\subsection*{Case $\muphi \neq 0$}

The case $\mu_\phi\neq0$ is a bit more subtle, because the mass matrix of $\h \Phi_0$ becomes $S$-dependent. Then a one-loop potential is generated,
and one has to check whether the two-loop runaway for $S$ dominates over the one-loop effects. The $\mu_\phi$ term in the superpotential of Eq.~\eqref{eq:Wtot} breaks explicitly the $U(1)'_A$ and $U(1)_R$ symmetries of the $\mu_\phi=0$ theory, and this will generate $\mu_\phi$-dependent masses for the axion fields corresponding to the phases of $X$ and $S$. The phase of $X$ receives its mass at tree-level, and it is given by Eq.~\eqref{eq:Xaxion}, while the phase of $S$ receives a mass from loop effects. Writing $S=S_0 \exp(i a_S/S_0)$, the one-loop masses for the modulus $S_0$ and the axion $\phi$ turn out to be
\begin{align}
\frac{ V^{(1)}}{16 \pi^2}\supset -\frac{h^2}{47\pi^2}\left(\frac{h\mu_\phi}{\mu_1}\right)^2 (h^2\muphi)^2 |S_0|^2
+
\frac{h^2}{48\pi^2}\Big(\frac{h\mu_\phi}{\mu_1}\Big)\Big(\frac{S_0}{\mu_1}\Big) (h^2\mu_\phi)^2\,a_S^2
+\ldots\,,
\label{eq:VCWmuphi1}
\end{align}
up to corrections that are higher order in $\muphi$; notice that there is no tadpole for $\phi$.

This has to be compared with the two-loop effects. The reason why two-loop effects can compete or dominate, is that the one-loop result (\ref{eq:VCWmuphi1}) is suppressed by additional powers of $\muphi$ as compared to the two-loop potential. Recalling that $\muphi$ is smaller than $\mu_1$, $\mu_2$, by a loop factor (see (\ref{eq:log-constraint})), the two-loop potential is in fact strictly larger than the one-loop effects. As explained before, this is due to the tree-level spectrum of the fields $\Phi_0$ and $\Psi_0$ becoming supersymmetric when $\muphi \to 0$.

The two-loop potential has a very complicated $\mu_\phi$ dependence, so that we will not provide complete analytical expressions but rather comment on its main features. It should be noted that the calculation of ref.~\cite{Giveon:2008wp} cannot be readily extended to the $\muphi\neq0$ case, since the $\mu_2=0$ theory still has a nonsupersymmetric spectrum, and the $\muphi$ interactions generate nonzero propagators between fermions of the same chirality, so that $V^{(2)}_{\overline{FF}S}$ in Eq.~\eqref{eq:VCW2} is not zero. The runaway for $S$  
is still generated, and agrees parametrically with Eqs.~\eqref{eq:Vsoftquad}~and~\eqref{eq:Vsoftlog}; this dominates over the one-loop result of~\eqref{eq:VCWmuphi1} (extended to the logarithmic regime). 

A similar effect is found for the axion, with the 2-loop $S$-axion mass dominating over the one-loop result. For the values of $S_0$ considered in the paper, the mass is of order
\begin{align*}
m^2_{a_S}\sim \frac{5\mu_2\mu_\phi}{(16\pi^2)^2}\,.
\end{align*}
Again there is no tadpole for $\phi$ at two loops.

\bibliographystyle{JHEP}

\begin{thebibliography}{10}

\bibitem{Dvali:1996cu}
  G.~R.~Dvali, G.~F.~Giudice, A.~Pomarol,
  ``The Mu problem in theories with gauge mediated supersymmetry breaking,''
  Nucl.\ Phys.\  {\bf B478}, 31-45 (1996).
  [hep-ph/9603238].

\bibitem{Komargodski:2008ax}
  Z.~Komargodski, N.~Seiberg,
  ``mu and General Gauge Mediation,''
  JHEP {\bf 0903}, 072 (2009).
  [arXiv:0812.3900 [hep-ph]].

\bibitem{Csaki:2008sr}
  C.~Csaki, A.~Falkowski, Y.~Nomura {\it et al.},
  ``New Approach to the mu-Bmu Problem of Gauge-Mediated Supersymmetry Breaking,''
  Phys.\ Rev.\ Lett.\  {\bf 102}, 111801 (2009).
  [arXiv:0809.4492 [hep-ph]].

\bibitem{SchaferNameki:2010iz}
  S.~Schafer-Nameki, C.~Tamarit, G.~Torroba,
  ``A Hybrid Higgs,''
  [arXiv:1005.0841 [hep-ph]].

\bibitem{Roy:2007nz}
  T.~S.~Roy, M.~Schmaltz,
  ``Hidden solution to the mu/Bmu problem in gauge mediation,''
  Phys.\ Rev.\  {\bf D77}, 095008 (2008).
  [arXiv:0708.3593 [hep-ph]].

\bibitem{Murayama:2007ge}
  H.~Murayama, Y.~Nomura, D.~Poland,
  ``More visible effects of the hidden sector,''
  Phys.\ Rev.\  {\bf D77}, 015005 (2008).
  [arXiv:0709.0775 [hep-ph]].

\bibitem{Hall:2002up}
  L.~J.~Hall, Y.~Nomura and A.~Pierce,
  ``R symmetry and the mu problem,''
  Phys.\ Lett.\  B {\bf 538}, 359 (2002)
  [arXiv:hep-ph/0204062].

\bibitem{Dine:2009swa}
  M.~Dine, J.~Kehayias,
  ``Discrete R Symmetries and Low Energy Supersymmetry,''
    [arXiv:0909.1615 [hep-ph]].

\bibitem{Delgado:2007rz}
  A.~Delgado, G.~F.~Giudice and P.~Slavich,
  ``Dynamical mu Term in Gauge Mediation,''
  Phys.\ Lett.\  B {\bf 653}, 424 (2007)
  [arXiv:0706.3873 [hep-ph]].



\bibitem{Dine:1993yw}
  M.~Dine and A.~E.~Nelson,
  Phys.\ Rev.\  D {\bf 48}, 1277 (1993)
  [arXiv:hep-ph/9303230].


\bibitem{Dine:1994vc}
  M.~Dine, A.~E.~Nelson and Y.~Shirman,
  ``Low-Energy Dynamical Supersymmetry Breaking Simplified,''
  Phys.\ Rev.\  D {\bf 51}, 1362 (1995)
  [arXiv:hep-ph/9408384].

\bibitem{Giudice:1997ni}
  G.~F.~Giudice and R.~Rattazzi,
  ``Extracting Supersymmetry-Breaking Effects from Wave-Function
  Renormalization,''
  Nucl.\ Phys.\  B {\bf 511}, 25 (1998)
  [arXiv:hep-ph/9706540].

\bibitem{Dine:2006xt}
  M.~Dine and J.~Mason,
  ``Gauge mediation in metastable vacua,''
  Phys.\ Rev.\  D {\bf 77}, 016005 (2008)
  [arXiv:hep-ph/0611312].

\bibitem{Dine:2007dz}
  M.~Dine and J.~D.~Mason,
  ``Dynamical Supersymmetry Breaking and Low Energy Gauge Mediation,''
  Phys.\ Rev.\  D {\bf 78}, 055013 (2008)
  [arXiv:0712.1355 [hep-ph]].

\bibitem{Giudice:2007ca}
  G.~F.~Giudice, H.~D.~Kim and R.~Rattazzi,
  ``Natural mu and Bmu in gauge mediation,''
  Phys.\ Lett.\  B {\bf 660}, 545 (2008)
  [arXiv:0711.4448 [hep-ph]].

\bibitem{Liu:2008pa}
  T.~Liu and C.~E.~M.~Wagner,
  ``Dynamically Solving the $\mu/B_\mu$ Problem in Gauge-mediated Supersymmetry
  JHEP {\bf 0806}, 073 (2008)
  [arXiv:0803.2895 [hep-ph]].

\bibitem{Mason:2009iq}
  J.~D.~Mason,
  ``Gauge Mediation with a small mu term and light squarks,''
  Phys.\ Rev.\  D {\bf 80}, 015026 (2009)
  [arXiv:0904.4485 [hep-ph]].

\bibitem{Evans:2010ru}
  J.~L.~Evans, M.~Sudano and T.~T.~Yanagida,
  ``A CP-safe solution of the mu/ Bmu problem of gauge mediation,''
  arXiv:1008.3165 [hep-ph].

\bibitem{Giudice:1998bp}
  G.~F.~Giudice and R.~Rattazzi,
  ``Theories with gauge-mediated supersymmetry breaking,''
  Phys.\ Rept.\  {\bf 322}, 419 (1999)
  [arXiv:hep-ph/9801271].

\bibitem{Luty:1998vr}
  M.~A.~Luty, J.~Terning,
  ``Improved single sector supersymmetry breaking,''
  Phys.\ Rev.\  {\bf D62}, 075006 (2000).
  [hep-ph/9812290].


\bibitem{Ellis:1988er}
  J.~R.~Ellis, J.~F.~Gunion, H.~E.~Haber {\it et al.},
  ``Higgs Bosons in a Nonminimal Supersymmetric Model,''
  Phys.\ Rev.\  {\bf D39}, 844 (1989).

\bibitem{deGouvea:1997cx}
  A.~de Gouvea, A.~Friedland and H.~Murayama,
  ``Next-to-minimal supersymmetric standard model with the gauge mediation  of
  supersymmetry breaking,''
  Phys.\ Rev.\  D {\bf 57}, 5676 (1998)
  [arXiv:hep-ph/9711264].



\bibitem{Dine:2006gm}
  M.~Dine, J.~L.~Feng, E.~Silverstein,
  ``Retrofitting O'Raifeartaigh models with dynamical scales,''
  Phys.\ Rev.\  {\bf D74}, 095012 (2006).
  [hep-th/0608159].

\bibitem{Essig:2007xk}
  R.~Essig, K.~Sinha, G.~Torroba,
  ``Meta-stable dynamical supersymmetry breaking near points of enhanced symmetry,''
  JHEP {\bf 0709}, 032 (2007).
  [arXiv:0707.0007 [hep-th]].

\bibitem{Dine:2007xi}
  M.~Dine, N.~Seiberg, S.~Thomas,
  ``Higgs physics as a window beyond the MSSM (BMSSM),''
  Phys.\ Rev.\  {\bf D76}, 095004 (2007).
  [arXiv:0707.0005 [hep-ph]].


\bibitem{Intriligator:2006dd}
  K.~A.~Intriligator, N.~Seiberg, D.~Shih,
  ``Dynamical SUSY breaking in meta-stable vacua,''
  JHEP {\bf 0604}, 021 (2006).
  [hep-th/0602239].

\bibitem{Franco:2006es}
  S.~Franco, A.~M .Uranga,
  JHEP {\bf 0606}, 031 (2006).
  [hep-th/0604136].

\bibitem{Giveon:2008wp}
  A.~Giveon, A.~Katz and Z.~Komargodski,
  ``On SQCD with massive and massless flavors,''
  JHEP {\bf 0806}, 003 (2008)
  [arXiv:0804.1805 [hep-th]].

\bibitem{Martin:2001vx}
  S.~P.~Martin,
  ``Two-loop effective potential for a general renormalizable theory and
  softly broken supersymmetry,''
  Phys.\ Rev.\  D {\bf 65}, 116003 (2002)
  [arXiv:hep-ph/0111209].

\bibitem{Intriligator:2008fe}
  K.~Intriligator, D.~Shih, M.~Sudano,
  ``Surveying Pseudomoduli: The Good, the Bad and the Incalculable,''
  JHEP {\bf 0903}, 106 (2009).
  [arXiv:0809.3981 [hep-th]].

\bibitem{Giveon:2008ne}
  A.~Giveon, A.~Katz, Z.~Komargodski, D.~Shih,
  ``Dynamical SUSY and R-symmetry breaking in SQCD with massive and massless flavors,''
  JHEP {\bf 0810}, 092 (2008).
  [arXiv:0808.2901 [hep-th]].

\bibitem{Essig:2008kz}
  R.~Essig, J.~-F.~Fortin, K.~Sinha, G~Torroba, M.~Strassler,
  ``Metastable supersymmetry breaking and multitrace deformations of SQCD,''
  JHEP {\bf 0903}, 043 (2009).
  [arXiv:0812.3213 [hep-th]].


\bibitem{Seiberg:1994pq}
  N.~Seiberg,
  ``Electric - magnetic duality in supersymmetric nonAbelian gauge theories,''
  Nucl.\ Phys.\  {\bf B435}, 129-146 (1995).
  [hep-th/9411149].

\bibitem{Craig:2009hf}
  N.~Craig, R.~Essig, S.~Franco, S.~Kachru, G.~Torroba,
  ``Dynamical Supersymmetry Breaking, with Flavor,''
  Phys.\ Rev.\  {\bf D81}, 075015 (2010).
  [arXiv:0911.2467 [hep-ph]].


\bibitem{Cheung:2010jx}
  C.~Cheung and Y.~Nomura,
  arXiv:1008.5153 [hep-ph].


\bibitem{Franco:2009wf}
  S.~Franco and S.~Kachru,
  ``Single-Sector Supersymmetry Breaking in Supersymmetric QCD,''
  arXiv:0907.2689 [hep-th].

\bibitem{Behbahani:2010wh}
  S.~R.~Behbahani, N.~Craig, G.~Torroba,
  ``Single-sector supersymmetry breaking, chirality, and unification,''
  [arXiv:1009.2088 [hep-ph]].

\bibitem{Ford:1992pn}
  C.~Ford, I.~Jack and D.~R.~T.~Jones,
  ``The Standard Model Effective Potential at Two Loops,''
  Nucl.\ Phys.\  B {\bf 387}, 373 (1992)
  [Erratum-ibid.\  B {\bf 504}, 551 (1997)]
  [arXiv:hep-ph/0111190].




\end{thebibliography}
\renewcommand{\refname}{Bibliography}
\addcontentsline{toc}{section}{Bibliography}
\providecommand{\href}[2]{#2}\begingroup\raggedright

\end{document}